\documentclass[conference,a4paper]{APSIPA2021}
\usepackage{multirow}
\usepackage{graphicx}
\usepackage{amsmath}
\usepackage{amsxtra}
\usepackage{threeparttable}
\usepackage{makecell, multirow, lineno, hyperref}

\usepackage{geometry}
\usepackage{fancyhdr}
\renewcommand{\headrulewidth}{0pt}
\renewcommand{\footrulewidth}{0pt}
\fancypagestyle{firststyle} {
    \fancyhf{}
    \fancyhead[L]{Proceedings of 2022 APSIPA Annual Summit and Conference}
    \fancyhead[R]{7-10 November 2022, Chiang Mai, Thailand}
    
    \fancyfoot[L]{978-616-590-477-3 ©2022 APSIPA}

    \fancyfoot[R]{APSIPA ASC 2022}
}
\fancypagestyle{fancy} {
    \fancyhf{}
    \fancyhead[L]{Proceedings of 2022 APSIPA Annual Summit and Conference}
    \fancyhead[R]{7-10 November 2022, Chiang Mai, Thailand}
    
}

\begin{document}

\title{Joint Speech Activity and Overlap Detection with Multi-Exit Architecture}

\author{%
\authorblockN{%
Ziqing Du, Kai Liu, Xucheng Wan, Huan Zhou
}
\authorblockA{%
Artificial Intelligence Application Research Center, Huawei Technologies \\Shenzhen, PRC\\
E-mail: {\{duziqing1, liukai89, wanxucheng, zhou.huan\}@huawei.com}}
}

\maketitle
\begin{abstract}
Overlapped speech detection (OSD) is critical for speech applications in scenario of multi-party conversion. Despite numerous research efforts and progresses, comparing with speech activity detection (VAD), OSD remains an open challenge and its overall performance is far from satisfactory. The majority of prior research typically formulates the OSD problem as a standard classification problem, to identify speech with binary (OSD) or three-class label (joint VAD and OSD) at frame level. In contrast to the mainstream, this study investigates the joint VAD and OSD task from a new perspective. In particular, we propose to extend traditional classification network with multi-exit architecture. Such an architecture empowers our system with unique capability to identify class using either low-level features from early exits or high-level features from last exit. In addition, two training schemes, knowledge distillation and dense connection, are adopted to further boost our system performance. Experimental results on benchmark datasets (AMI and DIHARD-III) validated the effectiveness and generality of our proposed system. Our ablations further reveal the complementary contribution of proposed schemes. With $F_1$ score of 0.792 on AMI and 0.625 on DIHARD-III, our proposed system outperforms several top performing models on these datasets, but also surpasses the current state-of-the-art by large margins across both datasets.  Besides the performance benefit, our proposed system offers another appealing potential for quality-complexity trade-offs, which is highly preferred for efficient OSD deployment. 
\end{abstract}


\section{Introduction}
The occurrence of multiple talkers speak simultaneously is common and natural in spontaneous human conversations, especially in scenario of multi-party meetings. The presence of overlapped speech segments however has adverse impacts on most speech analysis systems (such as speech recognition, speaker identification or diarization), which are typically designed in the absence of overlapped speech. For example, it has been reported \cite{shriberg2001observations} there is a significant increase in word error rate in segments containing overlapped speech (12\% absolute).

\subsection{Prior art}
To address the issue, various research approaches on overlapped speech detection (OSD) have been developed over the last decade. One popular research direction is to consider OSD as an independent front-end pre-processing task. It can be formulated as a classification problem, to identify each speech frame with binary pairwise or three-class label (non-speech, overlapped speech and single speaker speech). Following the pioneer work \cite{GeigerESR13} that illustrated the success of LSTM-based OSD system, different DNN architectures have been explored, together with various classification and feature engineering techniques. 

On top of LSTM-based OSD, Sajjan et al. \cite{SajjanGSGR18} further showed that LSTM-based models with spectrogram feature provided good separation between single speaker and overlap classes; and instead of using well established speech features, Bi-LSTM applied on trainable SincNet features \cite{BullockBG20,sinc2018}. Using CNN-based architecture, \cite{Andrei2017} investigated how frame duration influences the OSD accuracy; On basis of short frames, \cite{YousefiH20} explored the effects of different features; \cite{Kunesov2019} evaluated CNN-based OSD in terms of the potential improvements to speaker diarization. Instead of performing frame-level or fixed-length OSD, \cite{zhang21w} used deep feed-forward sequential memory networks to perform segment-level OSD by leveraging spatial information from multi-channel speech recordings. Using MFCC feature, architecture in \cite{RajHK21}, consisted of time-delay neural network layers and bidirectional LSTM layers, performed 3-class frame-level classification for overlap detection.

In parallel, another research line is to jointly optimize OSD with downstream task in an end-to-end (E2E) approach. For example, an E2E neural speaker diarization was recently presented in \cite{TakashimaF0HGN21}, which proposed a multitask learning framework that optimizes speaker diarization conditioned on voice activity detection (VAD) and OSD as two subtasks; by augmenting a RNN-T-based model, a multi-talker RNN-T \cite{Tripathi20} was designed to recognize speech with multiple talkers; to jointly perform OSD as well as speaker counting, an CNN-based framework was proposed in \cite{zhangwy19}, convolutional recurrent neural networks (CRNN) was used in \cite{countnet19} and Temporal Convolutional Network architecture was proposed for joint VAD, OSD and speaker counting tasks \cite{CornellOS020}.

\subsection{Existing problems}
Despite these efforts, handling overlapped speech is still challenging and remains an open problem, as claimed in the recent reports on DIHARD III \cite{ryant21} and CHIME-6 \cite{watanabe20b_chime}. Even with increased amount of overlapped speech from augmentation scheme, the identification of OSD is still much less reliable than that of VAD due to inherent the nature of short segment length and mixture  (e.g.\cite{Andrei2017,Kunesov2019,SajjanGSGR18, chen21t_interspeech, CornellOS020}).

\subsection{Multi-exit architecture}
In this paper, our focus is on joint VAD and OSD task. We note that most previous research works share two things in common: 1) classification features are extracted from the same representation space; and 2) classification networks are lightweight with only a few layers. Unlike these prior works, our intuition is that vast majority OSD frames are intrinsically more difficult to be identified than VAD frames. As such, intuitively, we hypothesize that not all speech frames require the same amount of network computation to yield a confident classification.  

Based on this hypothesis, we expect to design a network with capability to adapt classification features from different latent spaces for each input sample. To this end, we propose a joint VAD and OSD network based on multi-exit architecture (MEA), which is appealing for improving inference efficiency by predicting easy samples at early exits and hard samples at later exits. To the best of our knowledge, MEA is here explored for the first time on the OSD task. 

The remainder of this paper is structured as follows. More details on MEA are introduced in Section 2; our proposed OSD system is described in Section 3; Section 4 reports experimental results; and the paper is concluded in Section 5.

\section{Related Works}

Put simply, MEA is a layered classification architecture, augmented by early exits that are inserted after intermediate layers. It is regarded as a member of dynamic network family\cite{han21}, which is an emerging direction and has gained increasing attention on image classification.

MEA mainly adopts the early exiting method for dynamic inference, which allows prediction to quit the network early when samples can already be inferred with high confidence. The first work to propose attaching early exits to a deep network was \cite{Teerapittayanon16}, where standard image classification architectures such as LeNet, AlexNet and ResNet, were augmented by early exits. Later Huang et al.\cite{HuangCLWMW18} proposed Multi-Scale DenseNet, which was the state-of-the-art MEA. 

\begin{figure}[htb]
\begin{center}
\includegraphics[width=85mm]{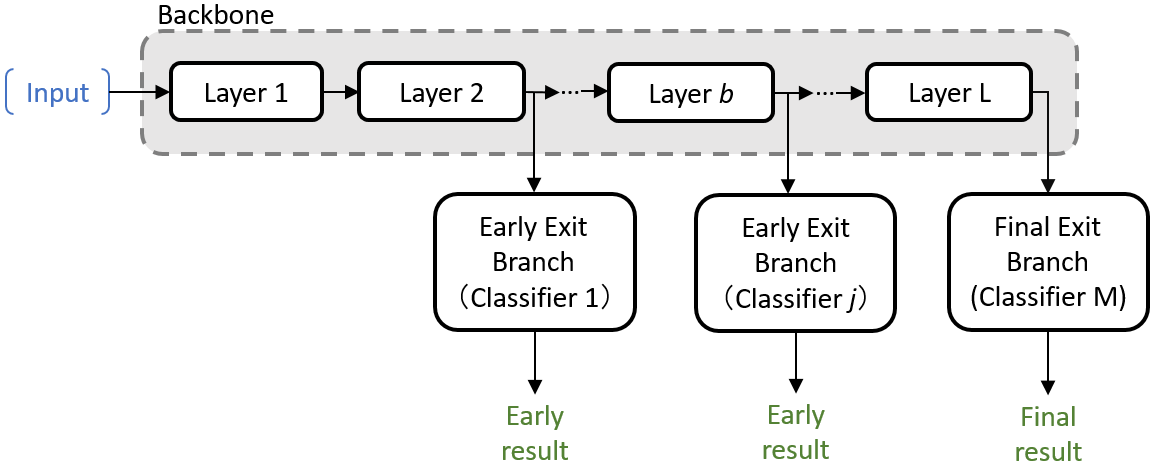}
\end{center}
\caption{Illustration of general multi-exit architecture}
\label{fig:mea}
\end{figure}

Fig.\ref{fig:mea} depicts general diagram of an MEA network with $M$ exits. The backbone network as shown in dashed box consists of $L$ layers which help introduce branch classifiers settled at different depths of the network. 
Thus a sequence of intermediate exits' classifier $(p_1, \dots, p_{M-1})$ along with final exit's classifier $p_{M}$ are formed. Clearly, classifier at later exit is more accurate and more expensive to compute than the previous classifier. The goal of the network is to learn $f: \mathcal{X} \rightarrow y$ that maps an input space $\mathcal{X}$ to a vector of class prediction scores $c = [c_1,\dots,c_K]^T$, where $c_i \in [0,1]$ and $K$ is the number of class. If  input $\mathcal{X}_i$ is predicted at the $m^{th}$ exit, its output can be expressed as $c^{(m)}_i = f(\mathcal{X}_i;\theta^m)$ where $c^{(m)}_i$ denotes the probability distribution over $K$ classes and $\theta^m$ refers to weights up to the $m^{th}$ exit.

MEAs are typically trained with a multi-task objective: one attaches a loss function to each exit, for example cross-entropy, and minimizes the sum of exit-wise losses, as if each exit formed a separate classification task. During inference, these exits are evaluated in turn and allow to terminate the inference procedure at an intermediate layer, called adaptive inference. Depending whether the computational budget is available, the inference of an multi-exit system can operate in either budget-mode or anytime-mode \cite{Phuong19}. 

Note that researches in MEA-based network mostly focused on its key advantage, that is, reducing computation and save energy using adaptive inference \cite{Teerapittayanon16, HuangCLWMW18}. In contrast, our main objective in this paper, is to classify easy samples at the earliest possible exit, as well as improve the classification performance on hard samples, through leveraging a MEA design during training.

\section{Proposed Methods}
\begin{figure}[htb]
\centering
\includegraphics[width=8.5cm]{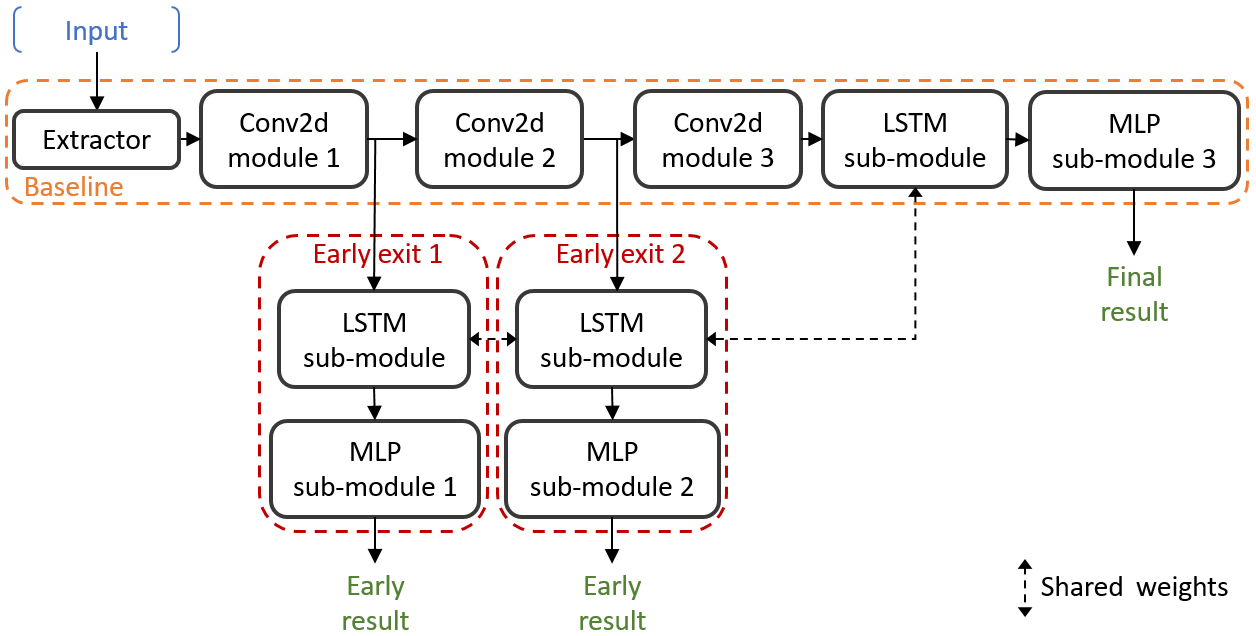}
\caption{Illustration of our proposed system with multi-exit architecture}
\label{fig:meaosd}
\end{figure}

The overall architecture of our proposed system is illustrated in Fig.\ref{fig:meaosd}. Its core modules which are built on the basis of CRNN network and training schemes are described respectively in the following subsections. 

\subsection{Overall Architecture}
Our proposed system consists of a baseline subnet and two early exit modules. The baseline subnet (i.e., the orange dashed box in Fig.\ref{fig:meaosd}) includes one extractor, three Conv2d modules (stacked\_Conv2d) and one exit module (composed of one LSTM and one MLP sub-module). The whole process can be outlined below: 
\begin{equation}
\label{eq:network}
\begin{aligned}
y^\prime &= Extractor(X)\\
y^{\prime\prime} &= stacked\_Conv2d(y^\prime)\\
\hat{y} &= Exit(y^{\prime\prime}) = MLP(LSTM(y^{\prime\prime}))\\
\end{aligned}
\end{equation}
To augment the baseline with MEA, two additional early exit modules are attached, placed after first two Conv2d modules respectively. To differentiate from the early exits, the original exit in the baseline is referred as the final exit afterwards. All module details are described below. 
\subsubsection{\textbf{Extractor}}
Our system takes raw speech waveform as the input. With motivation to build an adaptive front-end, the extractor module is constructed by SincNet \cite{sinc2018}, four 2D-convolution layers, two average pooling layers and squeeze-and-excitation (SE) layers \cite{se}. Here, instead of using traditional handcrafted speech features, we use the SincNet features with cut-off frequencies learned from the raw waveform. The feature dimension is further reduced through the subsequent layers by exploring temporal pattern among features. 

\subsubsection{\textbf{Conv2d module}}
The feature outputs from extractor are further mapped by three cascaded Conv2d modules. Each module includes two convolution layers with different configuration of filter number and kernel size. 
Note that the output dimension of each Conv2d module is consist to that of the input, to ease attachment of multiple exits with shared LSTM parameters.

\subsubsection{\textbf{LSTM sub-module}}
Inspired by the architecture of CRNN network \cite{crnn2015}, LSTM sub-module is placed right after those CNN modules to aggregate feature sequence. Each sub-module has one average pooling layer and one Bi-LSTM layer. It is important to note that, to reduce the capacity of the model, these LSTM sub-modules share their parameters across exits.

\subsubsection{\textbf{MLP sub-module (Classification)}}
Each MLP sub-module includes two fully connected layers. The former is employed for dimension reduction while the latter noted as classifier for classification. For objective of joint VAD and OSD, this module predicts the probability of occurrence of three classes, labeled as \{0: non-speech, 1: single speaker speech, 2: overlapped speech\}. 

Last of all, we point out that above proposed system architecture is general and can be easily extended to handle other multi-class prediction problems (say, joint activity detection and concurrent speaker counting). 



\subsection{Training and Inference Schemes}
As aforementioned, training an MEA is not trivial. Specific training techniques are needed to address our unique optimization objective. Due to lack of prior information on input complexity, our training principle is to train all exits with objective of overall classification quality. 

Nevertheless, jointly optimizing all classifiers within a chain-structured classification system is challenging. In order to improve the back propagation of gradient and make the network easier to train, different feature combination utilization schemes are considered to generate more accurate training supervision. We propose to deploy two possible schemes, knowledge distillation (KD) and dense connection (DC).

\subsubsection{\textbf{KD-based loss}}
The motivation behind the scheme is to introduce a possibility for different exits to learn from each other. As widely known, KD is a popular technique for knowledge transfer from a pre-trained large-sized network (teacher) to a small-sized one (student). For case of training MEA, we assume that an ensemble representation, as a summary statistic of individual representations from multiple exits, can play a role of teacher. It triggers us to adopt a KD-based method as in \cite{lee21}, which transfers knowledge by encouraging every exit to mimic the probabilistic outputs of the teacher exit. 
\begin{figure}[htb]
\centering
\includegraphics[width=5.5cm]{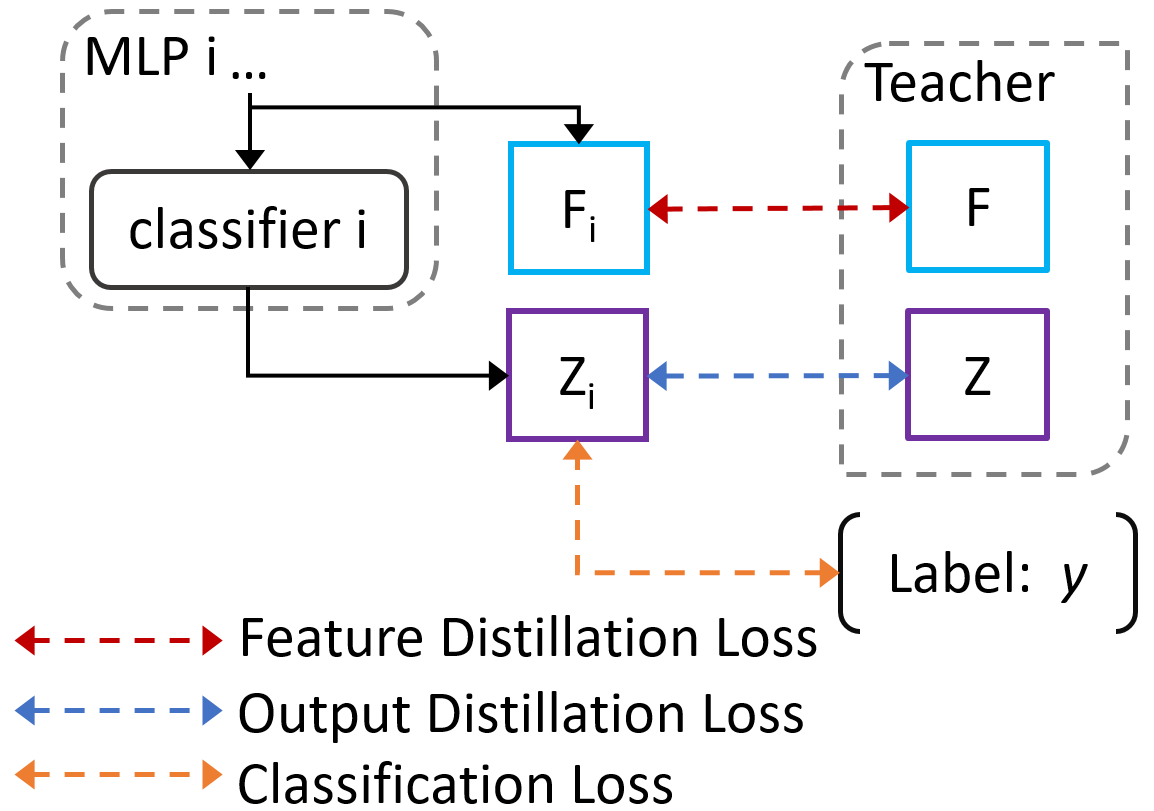}
\caption{Illustration of the knowledge distillation based training strategies at an exit.}
\label{fig:kd}
\end{figure}

Guided by the idea, we introduce two kinds of KD-losses with respective teacher-student options (as illustrated in Fig.\ref{fig:kd}). In particular, for student option, we consider representations before or after each exit classifier, denoted by $F_i$ (latent features) and $Z_i$  (un-normalized probability outputs) for the $i$th-exit respectively. As for teacher $F$ and $Z$, we adopt the ensemble information from all exits, which can be simplified as $Z =\frac{{1}}{{M}}\sum^{M}_{i=1} Z_i $ and $F =\frac{{1}}{{M}}\sum^{M}_{i=1} F_i $.

With the KD-based loss, the overall training loss $L_{joint}$ can be expressed as a joint loss. It combines an aggregated conventional classification loss $L_C$ with a probability-based distillation loss $L_Z$ and a feature-based distillation loss $L_F$, weighted by coefficients $\alpha$ and $\beta$. That is, \begin{equation}
\label{eq:1}
\begin{aligned}
L_{joint} &= L_C + \alpha L_Z + \beta L_F \\
&= \sum^{M}_{i=1} \Big[L_C(Z_i,y) + \alpha L_Z(Z_i,Z)+ \beta L_F(F_i,F)\Big] \\
\end{aligned}
\end{equation}
Note that the additional two KD-loss items do not require access to the actual ground-truth $y$.

\subsubsection{\textbf{DC-based scheme}}
Alternatively, another training strategy is to enforce the network to explicitly control the flow of information via feature fusion. In particular we adopt the DC scheme \cite{HuangCLWMW18}, which uses dense inter-connections to concatenate previous features from early backbone stages to features at later stages. Note that different from the skip connection scheme, DC scheme ensures feature reusability, where output feature of one layer is concatenated to the input to the next layers rather than a summation.

\begin{figure}[htb]
\centering
\includegraphics[width=5.3cm]{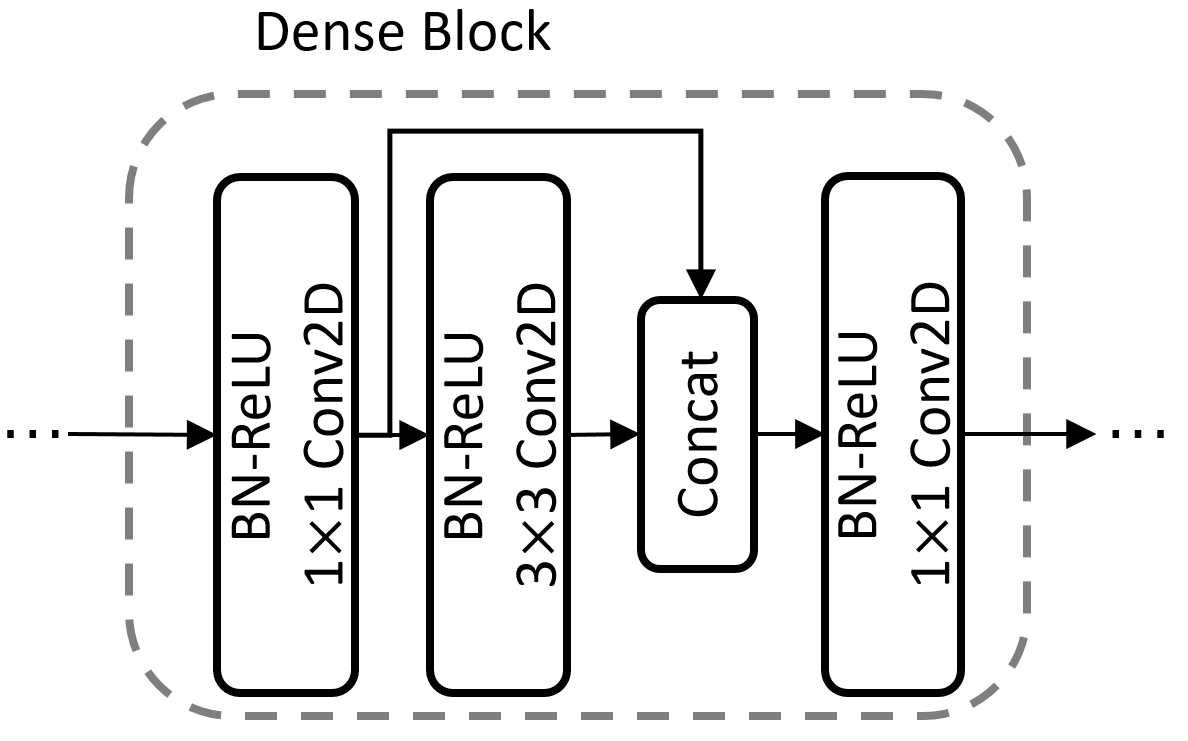}
\caption{The subnet module supporting DC scheme}
\label{fig:dense}
\end{figure}

To accommodate this training scheme, the original Conv2d module introduced earlier is accordingly modified to a dense block (as illustrated in Fig.\ref{fig:dense}). The yielded new module is called as \textit{Conv2d-DC}. 


\subsubsection{\textbf{Inference mode}}
Recall that our main objective is twofold: to achieve overall decent classification performance and to verify the possibility of dynamic inference with computation-accuracy tradeoffs. To this end, we propose two inference modes to access each objective individually. 

The first inference mode, called normal mode, is to treat our proposed system as a conventional classification system, by only examine system outputs at the final exit. The other mode is to evaluate all exit outputs under a given threshold $\gamma$. That is, for a test sample, its prediction probabilities at all exits are collected in order. If any of them is higher than the given threshold, the corresponding class label is assigned to the sample, otherwise, the class label at the final exit is used instead. This inference strategy is termed exiting mode. 



\section{Experiments}
\subsection{Experimental Setup}
\subsubsection{\textbf{Datasets}}
Three benchmark datasets are utilized for our system training: AMI Meeting \cite{carletta2007unleashing}, DIHARD III \cite{ryant21} and VoxConverse dataset \cite{chung2020spot}. Apart from the officially designated training dataset, we also generate synthesized data by weighted downmixing of two random speech samples, which offers additional 40\% training data. In addition, following a common practice, speech samples are mixed on-the-fly during the training process, which augments each training sample with random noise and reverberation component based on MUSAN and RIRS\_NOISES \cite{snyder2015musan,ko2017study} respectively. For system evaluation, both test sets in the DIHARD III and AMI are used to examine our system on both performance and generality.
In such a setting, a summary of dataset statistics is reported in Table \ref{tab:res}, where training data comprises both official data and our synthesized data. In addition, for the test datasets, the class ratios of both VAD and OSD are provided. Note that the imbalance ratio (IR) between VAD-class and OSD-class is severe, up to 1:11.3, on the DIHARD III test set. 

\begin{table}[htb]
  \renewcommand\arraystretch{1.15}
  \caption{Dataset statistics }
  \label{tab:res}
  \centering
  \begin{tabular}{l|ccc|cccc}
\hline
\hline
\multirow{2}{*}{Dataset}  & \multicolumn{3}{c}{Duration (h)}\vline
& \multicolumn{2}{c}{Class Percentage (\%)}\\
    & Train& Dev.& Test & OSD & VAD\\
\hline
  DIHARD III & 25 + 10 & 9 & 33 &7 &  79\\
      AMI & 22 + 8.5  & 11  & 9 & 16& 83\\
  VoxConverse & 15 + 6 & 6  & - & - & - \\
\hline
\hline
  \end{tabular}
\end{table}

\subsubsection{\textbf{Implementation details}}
Our model takes 1.5s monophonic speech chunks (with sampling rate of 16kHz) as input, and outputs 50 class labels per speech chunk. This corresponds to generate classification results every 30ms. 

Full implementation details of our system can be found in Table \ref{tab:ressett}, along with the count (M) of parameters of core modules. The total number of parameters of the model is up to 1.5M when the DC-based scheme is applied (otherwise, it's around 1.3M).

\begin{table}[htb]
  \setlength\tabcolsep{4pt}
  \renewcommand\arraystretch{1.15}
  \caption{Implementation details of our proposed system}
  \label{tab:ressett}
  \centering
  \begin{tabular}{c|c|c|c}
\hline
\hline
Module & Setting & Output shape& \#Parameters\\
\hline
Input & - & $1\times24000$& - \\
\hline
\multirow{5}{*}{Extractor}&  SincNet(128,251,80) & \multirow{5}{*}{$64\times32\times50$}&\multirow{5}{*}{0.1}  \\
& Conv2d(32,3,1)$\times2$&\\
& SE(4) + Avg\_pool(2,1)&\\
& Conv2d(64,3,1)$\times2$ &\\
& SE(4) + Avg\_pool(3,2)&\\
\hline
Conv2d & Conv2d(256,1,1)&\multirow{2}{*}{$64\times32\times50$} &\multirow{2}{*}{0.5} \\
 ($\times3$) & Conv2d(64,3,1)& \\
\hline
\multirow{3}{*}{ \shortstack{Conv2d-DC \\($\times3$) }}&  Conv2d(320,1,1) &\multirow{3}{*}{$64\times32\times50$}&\multirow{3}{*}{0.7} \\
  & Conv2d(80,3,1)&\\
  &  Conv2d(64,1,1)&\\
\hline
\multirow{2}{*}{LSTM}& Avg\_pool(32) + Permute&\multirow{2}{*}{$50\times256$}&\multirow{2}{*}{0.6} \\
& Bi-LSTM(128)$\times2$&\\
\hline
MLP & Linear(128) &\multirow{2}{*}{$50\times3$}&\multirow{2}{*}{0.1}\\
 ($\times3$) & Linear(3)& \\
\hline
\hline
  \end{tabular}
\end{table}
Our system is optimized using the training loss $L_{joint}$, defined in (\ref{eq:1}). Here we adopt the proportionally-weighted cross entropy (CE) loss for $L_c$ and the Kullback-Leibler (KL) divergence loss for $L_z$ and $L_F$, respectively. In addition, weights are set by $\alpha=0.5$ and $\beta=1.0$. With a batch size of 256, our system is trained for 50 epochs using Adam optimizer. The initial learning rate is 0.001 and scales with a factor of 0.6 when there is no loss decrease over 6 epochs.

\begin{table*}[htb]
  \renewcommand\arraystretch{1.15}
  \caption{Performance comparisons on the VAD task (symbol $\downarrow$ indicates lower value represents better performance)}
  \label{tab:vad}
  \centering
  \begin{tabular}{l|ccc|ccc}
    \hline
    \hline
    
    VAD & \multicolumn{3}{c}{\makecell{AMI}} \vline& \multicolumn{3}{c}{\makecell{DIHARD III}}  \\
    
    System &FA $\downarrow$&Miss $\downarrow$ &ER $\downarrow$&FA $\downarrow$ &Miss $\downarrow$ &ER $\downarrow$ \\
    \hline 
    silero vad \cite{SileroVAD} &9.4& \textbf{1.7}&11.0&17.0&4.0&21.0\\
    pyannote 1.1 \cite{pyau11} &6.5& \textbf{1.7}&8.2&4.1&3.8&7.9\\
    pyannote 2.0 \cite{Bredin2021}&3.6&3.2&6.8&3.9&\textbf{3.3}&7.3\\
    Ours& \textbf{1.3}& 5.0 & \textbf{6.3} & \textbf{2.0} &5.0 &\textbf{7.0} \\
    \hline
    \hline
  \end{tabular}
\end{table*}

\begin{table*}[htb]
  \renewcommand\arraystretch{1.15}
  \caption{Performance comparisons on the OSD task (symbol $\uparrow$ means the higher value, the better)}
  \label{tab:osd}
  \centering
  \begin{tabular}{l|ccc|ccc|ccc|ccc}
    \hline
    \hline
    OSD & \multicolumn{6}{c}{\makecell{AMI}}\vline & \multicolumn{6}{c}{\makecell{DIHARD III}}  \\
    System & FA $\downarrow$ & Miss $\downarrow$ & ER $\downarrow$ & Precision $\uparrow$ & Recall $\uparrow$ & $F_1$ $\uparrow$& FA $\downarrow$ & Miss $\downarrow$ & ER $\downarrow$ & Precision $\uparrow$ & Recall $\uparrow$& $F_1$ $\uparrow$\\
    \hline 
    pyannote 1.1 & 51.1&\textbf{12.1}& 63.2 &63.2&\textbf{87.9}&73.5&48.2&45.2& 93.4&53.2&54.8&54.0\\
    \textbf{Raj} et al. \cite{RajHK21} &- &- &- &\textbf{86.4} &65.2&74.3&-&-&-&-&-&-\\
    pyannote 2.0&\textbf{16.9} &29.4&46.3 &80.7& 70.5& 75.3& 46.9& 37.2& 84.1& 57.2& 62.8& 59.9\\
    Ours & 18.6 &22.3& \textbf{40.9} &80.7 &77.7 &\textbf{79.2} & \textbf{44.0}& \textbf{34.5}& \textbf{78.5} & \textbf{59.8} &\textbf{65.5}&\textbf{62.5 }\\
    \hline
    \hline
  \end{tabular}
\end{table*}

\subsubsection{\textbf{Inference details}}
Following our training configuration, our system yields up to 5 prediction labels per frame(30ms). Such multiple label options are further post-processed by applying the majority rule voting strategy. Then the one with a majority is chosen as the final predicted label.  

As a follow up to label predictions, VAD and OSD segmentation can be simply inferred, where the OSD segments are formed by concatenating the frames with class \{2: overlapped\}; and VAD segments are obtained by combining the frames of both classes \{1: single speaker, 2: overlapped speech\}.

\subsubsection{\textbf{Evaluation metrics}}
Inherited from prior studies, two sets of standard evaluation metrics are adopted in this study. One set of them includes the false alarm rate (FA) missed alarm rate (Miss) and detection error rate (ER); the other contains precision, recall and $F_1$-score ($F_1$). 

\subsection{Performance at Final Exit}
In this subsection, we evaluate our proposed system by checking its performance at the final exit. It includes the overall performance analysis and ablation studies on individual performance contribution from each proposed scheme. 

\subsubsection{\textbf{Overall Performance Analysis}}
To better evaluate the ability of our system for joint VAD and OSD, we expand our system evaluation by benchmarking it against several top performing prior arts. For a fair comparison, those competitive prior studies are published recently for OSD on the either AMI or DIHARD III dataset, and results reported in their original publications are quoted.   

All detailed comparison results are listed in Table \ref{tab:vad} and Table \ref{tab:osd} for VAD and OSD, respectively, where the best performance is bold-faced. 

Regarding performance on the VAD task, Table \ref{tab:vad} shows that in terms of error rate, our model performs best on both datasets. It considerately outperforms the previous SOTA, \textit{pyannote 2.0}, with relative improvement of 7.3\% and 4.1\% on the dataset AMI and DIHARD III, respectively. Besides, the table also reveals that: 1) the VAD system \cite{SileroVAD} has high false alarm rate; 2) \textit{pyannote 2.0} \cite{SileroVAD} outperforms its early version \cite{pyau11}; 3) our system shows the lowest false alarm rates and the best overall error rates. Also, it is worth noting that \textit{pyannote 2.0} \cite{SileroVAD} assigns specific hyper-parameters for different dataset, which is not realistic in practical scenario.  

A similar behavior is observed for performances on the OSD task. As shown in Table \ref{tab:osd}, our proposed system achieves more prominent performance gains over the existing systems. In particular, regarding $F_1$ score, our system surpasses \textit{pyannote 2.0} by 5.2\% relative on the AMI and 4.3\% on the DIHARD III. Notably, this advantage is more pronounced when measure with error rate, with relative 11.7\% improvement on the AMI and 6.7\% on the DIHARD III, respectively. In addition, on the more challenging DIHARD III test set, our system is considerably advantageous by outperforming prior studies over all metrics.   

To ours best knowledge, the results of \textit{pyannote\,2.0} are the best in the existing studies. By outperforming the \textit{pyannote\,2.0}, our system provides the state-of-the-art performance. In addition, it is worth noting that our system achieves such performance with even smaller model size (around 1.4 million parameters, comparing to 1.5 million used in \textit{pyannote\,2.0}). Lastly, again, the \textit{pyannote\,2.0} requires to determine hyper-parameters by cross-validation on a validation set, which limits its generalization. In contrast, our system use common parameters for all datasets and avoid the unrealistic dataset-specific tuning as much as possible.

\subsubsection{\textbf{Ablation Studies}} To examine individual impact of each proposed scheme, we conduct ablation studies on our system by disabling one scheme at a time. The according experimental results (in terms of $F_1$-scores) are reported in Table \ref{tab:resdh3}.

\begin{table}[htb]
  \renewcommand\arraystretch{1.15}
  \caption{Ablation study results in terms of $F_1$-score (percentage)}
  \label{tab:resdh3}
  \centering
  \begin{tabular}{l|cc|cc}
    \hline
    \hline
   \multirow{2}{*}{Model} & \multicolumn{2}{c}{\makecell{AMI}}\vline& \multicolumn{2}{c}{\makecell{DIHARD III}}\\
       &  \multicolumn{1}{c}{OSD} &\multicolumn{1}{c}{VAD} \vline & \multicolumn{1}{c}{OSD} &\multicolumn{1}{c}{VAD}\\ 

    \hline
    FULL   &79.2 &96.8  &62.5 &96.5 \\
    - DC-based scheme  &79.2 &96.9 &62.0 &96.6\\
    \quad- KD-based loss  & 79.0 & 96.7& 61.6 & 96.5\\
    \qquad- MEA (Baseline) & 77.2& 96.8& 61.6& 96.6\\
    \hline
    \hline
  \end{tabular}
\end{table}

From the table, we make the following key observations:
\begin{itemize}
    \item high VAD performance is maintained in all ablation experiments, which suggests our system has a very strong baseline; 
   \item disabling the DC-based scheme leads to 0.5\% absolute OSD performance drop on the DIHARD III test set;
   \item further disabling the KD-based scheme results in another 0.4\% absolute OSD performance drop on the DIHARD III test set;
    \item lastly, our baseline system by removing the MEA gets 1.8\% OSD performance degradation on the AMI test set, which indicates that MEA is the most important factor for our performance gains.
\end{itemize}
In all, by breaking down the performance gains on OSD task, it is proven that all our proposed schemes are beneficial and offer complementary contributions to the system performance. Additionally, among three schemes, our proposed architecture is dominant that boosts system performance the most.


\subsection{Performance at All Exits}
Recall that with leverage of MEA, our system provides a sequence of intermediate exits' classifier along with final exit's classifier. Thus far we have discussed the performance of final exit's classifier. Now we are interested in investigating those intermediate classifiers, by changing the inference mode to the exiting mode. 

In particular, since adaptive inference with reduced computation is beyond the scope of this study, our main objective herein is to investigate the feasibility of dynamic inference with computation-accuracy trade-offs. To verify the feasibility, our strategy is to analyze the exiting rate at all exits, given a high exiting threshold for the early exits. In this sense, we argue that if the proportions of early-exited samples, while small, are not negligible, then it becomes feasible to perform dynamic inference.

Motivated by the idea, in our experiments, a high threshold with value of 0.9 is set for those intermediate classifier at exit 1 and 2. Under such a setting, our inference process is switched from previous normal mode to the exiting mode. That is, test samples are firstly classified by the classifier at exit 1, those samples with high probability (no less than 0.9) are exited; the remaining samples are further classified by the classifier at exit 2, again, samples with high probability are exited; all remaining samples are classified by the last classifier at final exit. 

\begin{figure}[htb]
\centering
\includegraphics[width=8.3cm]{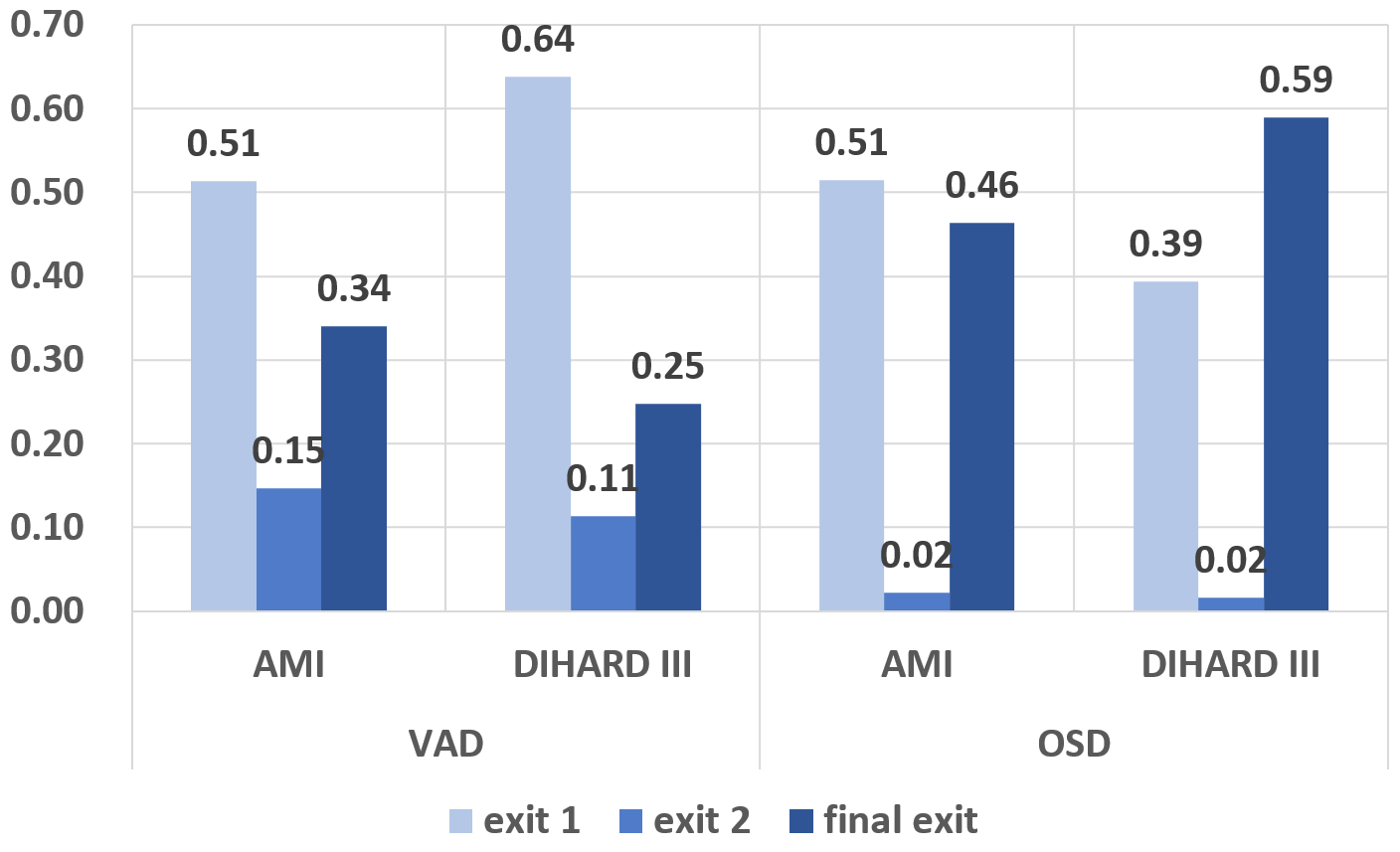}
\caption{Comparison of class distributions among exits}
\label{fig:per}
\end{figure}

With the said sequential inference process, the performance at all exists are evaluated. The resulting experimental results are illustrated in bar plots, as shown in Fig.\ref{fig:per}. Here the Y-axis denotes the exiting rate, which is calculated as percentage ratio of the number of samples exited at a specific exit to the total sample number, for a given class. In another words, a cluster with three adjacent bars reflects the class distribution among three exists.

From the bar plots, we can obtain a few valuable insights. Firstly, regarding the VAD task, our insights are:
\begin{enumerate}
\item similar distribution trend can be observed for both AMI and DIHARD III dataset. That is, the most samples are classified at exit 1 and the least samples at exit 2. It reveals the possibility to perform VAD classification with the dynamic inference, an interesting insight that has not be explored in previous studies; 
\item at least 51\% speech samples are easy that can be classified at exit 1. This implies the VAD detection may greatly benefited from the dynamic inference; 
\item about a quarter to one third samples are difficult that cannot be classified at early exits with high confidence.
\end{enumerate}
Secondly, regarding the OSD task, we observe that:
\begin{enumerate}
\item most samples are classified at exit 1 and final exit. And at least 41\% overlapped speech samples are easy that can be classified early, showing feasible possibility to perform dynamic inference;
\item more samples of DIHARD III are left to the final classifier for OSD judgment. This implies that on the OSD task, the DIHARD III dataset is more challenging than the AMI.
\end{enumerate}
Lastly, a close look on both tasks shows that, except the OSD task on the DIHARD III, over half samples can be classified at the earliest exit with the lowest complexity. Another finding is that both AMI and DIHARD III show higher OSD exiting rate at late exits (0.46 vs. 0.34 and 0.59 vs.0.25), which means OSD has inherently higher complexity than VAD.

Above results clearly confirm that variant sample complexities exist in both VAD and OSD class, which corroborate our motivation for adopting MEA to address the joint VAD and OSD problem. In addition, observations above suggest a large number of samples can be reliably classified by earlier exits that have lower capacity and complexity. This is a good match for objective of the dynamic inference. 

\section{Conclusions}
In this paper, we investigate the joint VAD and OSD problem, to classify speech frames into classes of non-speech, single speaker and overlapped speakers. Unlike any prior works, we study the problem from a new perspective by proposing a classification system with multi-exit architecture. Our design objective is twofold: improving the overall classification performance and classifying easy samples at the earliest possible exit. To boost system effectiveness, we also propose two training schemes to enhance the proposed architecture.  

To verify the efficacy of our proposed system, we conduct extensive experiments on the benchmark dataset of AMI and DIHARD III. Corresponding experimental results show that our system outperforms the previous top performing systems by a considerable margin. With $F_1$ score of 0.792 on AMI and 0.625 on DIHARD III, our system offers (to our best knowledge) the best OSD results reported till date. Beyond that, both training schemes are also proven to be effective and complementary via ablations. Meanwhile, our preliminary study shows that at least 41\% overlapped samples can be predicted at early exits with decent accuracy. This indicates our system also has the promising potential to balance performance and computational complexity, which would be further investigated in our future research. 



\end{document}